\documentclass{aa}  
\usepackage[varg]{txfonts}           

\usepackage{natbib}
\bibpunct{(}{)}{;}{a}{}{,} 
\usepackage{graphicx}

\usepackage[colorlinks=true, linkcolor=blue, citecolor=blue, urlcolor=blue]{hyperref}

\usepackage{placeins}

\usepackage{subfig}
\usepackage{subcaption}

\begin{document}

\title{The Sun at millimeter wavelengths}
\subtitle{V. Magnetohydrodynamic waves in a fibrillar structure}

\author{M.~Saberi\inst{1,2}, S.~Jafarzadeh\inst{3,4}, S.~Wedemeyer\inst{1,2}, R.~Gafeira\inst{5,6}, M.~Szydlarski\inst{1,2}, D. B. Jess\inst{7,8}, M. Stangalini\inst{9}}

\institute{Rosseland Centre for Solar Physics, University of Oslo, P.O. Box 1029 Blindern, 0315 Oslo, Norway\\
\email{maryam.saberi@astro.uio.no}
\and
Institute of Theoretical Astrophysics, University of Oslo, P.O. Box 1029 Blindern, 0315 Oslo, Norway
\and
Max Planck Institute for Solar System Research, Justus-von-Liebig-Weg 3, 37077 G\"{o}ttingen, Germany
\and
Niels Bohr International Academy, Niels Bohr Institute, Blegdamsvej 17, DK-2100 Copenhagen, Denmark
\and
Geophysical and Astronomical Observatory, Faculty of Science and Technology, University of Coimbra, Portugal
\and
Instituto de Astrofísica e Ciências do Espaço, Department of Physics, University of Coimbra, Portugal
\and
Astrophysics Research Centre, School of Mathematics and Physics, Queen’s University Belfast, Belfast, BT7 1NN, UK
\and
Department of Physics and Astronomy, California State University Northridge, Northridge, CA 91330, USA
\and
ASI Italian Space Agency, Via del Politecnico snc, I-00133 Rome, Italy
 }
\date{}

\abstract
{Magnetohydrodynamic (MHD) waves, playing a crucial role in transporting energy through the solar atmosphere, manifest in various chromospheric structures. Here, we investigated MHD waves in a long-lasting dark fibril using high-temporal-resolution (2~s cadence) Atacama Large Millimeter/submillimeter Array (ALMA) observations in Band 6 (centered at 1.25~mm). We detected oscillations in brightness temperature, horizontal displacement, and width at multiple locations along the fibril, with median periods and standard deviations of $240\pm114$~s, $225\pm102$~s, and $272\pm118$~s, respectively. Wavelet analysis revealed a combination of standing and propagating waves, suggesting the presence of both MHD kink and sausage modes. Less dominant than standing waves, oppositely propagating waves exhibit phase speeds (median and standard deviation of distributions) of $74\pm204$~km/s, $52\pm197$~km/s, and $28\pm254$~km/s for the three observables, respectively. This work demonstrates ALMA's capability to effectively sample dynamic fibrillar structures, despite previous doubts, and provides valuable insights into wave dynamics in the upper chromosphere.
}


\titlerunning{Magnetohydrodynamic waves in chromospheric fibrillar structures}
\authorrunning{M. Saberi et al.}
\maketitle

\section{Introduction}\label{Introduction}

The solar chromosphere is a highly dynamic environment featuring a wide range of magnetohydrodynamic (MHD) waves, which play a crucial role in transporting energy throughout the atmosphere \citep{2015SSRv..190..103J, 2016GMS...216..431V}. These waves hold the key to understanding the enigmatic heating of the Sun's outer layers \citep{1991mcch.conf.....U, 1993SoPh..143...49C}, the plasma-composition characteristics throughout the solar atmosphere \citep{2021ApJ...907...16B, 2021RSPTA.37900216S, 2021A&A...656A..87M, 2024PhRvL.132u5201M}, and the acceleration of solar wind \citep{1982SSRv...33..161L,2015NatCo...6.5947B}.
These waves are often generated in the low photosphere, as a result of interactions between plasma and magnetic fields, including processes like buffeting and interplay of magnetic flux tubes with surrounding granules \citep{1982SoPh...75....3S,1993SSRv...63....1S,2003A&A...406..725M} or magnetic reconnection events \citep{2009ApJ...705L.217H}. They propagate along magnetic-field lines throughout the atmosphere \citep[see, e.g.,][]{2011A&A...534A..65S}. Additionally, mode conversion may occur at the equipartition layers, where the sound and Alfv{\'e}n speeds nearly coincide \citep{2003ApJ...599..626B,2012A&A...538A..79N,2018NatPh..14..480G}. These processes contribute to the rich diversity of observable MHD wave types in the solar chromosphere, such as kink and sausage modes, which are linked to fluctuations in the transverse motion, width, and brightness of concentrated magnetic structures such as fibrils, spicules, and pores \citep[e.g.,][]{Edwin83, 2015ApJ...806..132G, 2022ApJ...930..129B}.

Fibrillar structures are ubiquitous features in the solar chromosphere, observed as dark or bright elongated features in various diagnostics, serving as potential conduits for wave propagation \citep{2012NatCo...3.1315M,2012ApJ...744L...5J,2017ApJS..229....9J,2017ApJS..229....7G,2021RSPTA.37900183M, 2024arXiv240515584B}. They are thought to represent bundles of concentrated magnetic field lines, manifested as dark or bright thread-like structures in intensity images, increasing their inclination angle (with respect to the solar normal) with height in the atmosphere as a result of the topology of the magnetic canopy \citep{1976RSPTA.281..339G,1982SoPh...79..267G, 1991A&A...250..220S,2009SSRv..144..317W}, whose heights depend on the magnetic-field strength at their footpoints \citep{2017ApJS..229...11J}. Oscillations in these structures can reveal the nature of various MHD wave modes, with periods typically ranging from a few minutes to tens of minutes \citep{2023LRSP...20....1J}. 

Accurately identifying MHD wave modes in magnetic structures like fibrils is important for estimating the energy each mode carries \citep{2023LRSP...20....1J}. Understanding wave energy dissipation is, in turn, crucial to unravelling the atmospheric energy budget \citep{2007Sci...318.1574D,2011Natur.475..477M}. While estimating the wave-energy flux based on observations has been possible, direct detection of energy deposition outside of sunspots and pores remains difficult \citep{2020ApJ...892...49H,2021RSPTA.37900172G,2021A&A...648A..77R}. Hence, wave energy is often estimated as a potential contributor to the overall atmospheric heating budget (i.e., if the wave energy is released in the solar chromosphere and/or corona), although direct evidence of its resulting thermalisation properties is often elusive.

The Atacama Large Millimeter/submillimeter Array (ALMA; \citealt{2009IEEEP..97.1463W}) provides a significant advancement in solar observations \citep[see, e.g.,][]{2016SSRv..200....1W,2017SoPh..292...87S,2017SoPh..292...88W,2019AdSpR..63.1396L,2022FrASS...981205N}. Its high-temporal resolution ($1-2$~s cadence) enables the study of high-frequency waves in the solar chromosphere \citep{2021RSPTA.37900184G,2022A&A...665L...2G,2023A&A...671A..69G}, unconstrained by the current spatial-resolution limitation which can also affect the detection of high frequencies \citep{2021A&A...656A..68E}. Moreover, we can measure brightness temperature with ALMA rather than intensity (common in other non-millimetre observations; \citealp[e.g.,][]{2019A&A...622A.150J}). This provides a direct connection to gas temperatures, offering new insights into wave dynamics and energy deposition. Thus, ALMA acts as a linear thermometer, directly measuring chromospheric heating signatures \citep{2016SSRv..200....1W, 2020A&A...635A..71W}. Beyond advances in wave studies \citep{2020A&A...634A..86P,2021A&A...652A..92N,2021RSPTA.37900174J,2022ApJ...924..100C}, ALMA's capabilities extend to analysing shock phenomena \citep{2021RSPTA.37900185E,2021ApJ...906...83C}, small-scale dynamic events \citep{2020A&A...644A.152E}, and the response of chromospheric structures to transient events \citep{2020A&A...643A..41D,2020A&A...638A..62N}, further broadening our understanding of solar atmospheric dynamics. Similarities between ALMA Band 3 (centered at 3 mm) brightness temperature maps and H$\alpha$ line-width images have been reported by \citet{2019ApJ...881...99M}.

In this paper, we investigate oscillations in brightness temperature, width, and transverse displacement of a well-identified dark fibrillar structure observed with ALMA in Band 6 (centered at 1.2 mm). Such dark fibrillar structures, manifesting within intensity images in the mid-to-upper chromospheric magnetic canopy, are regularly observed in other diagnostics, such as H$\alpha$ (see, e.g., \citealt{2009ApJ...705..272R, 2012NatCo...3.1315M, 2024arXiv240515584B}). 
However, \citet{2017A&A...598A..89R} predicted that while they would also be readily apparent in ALMA millimeter observations, exhibiting a similar appearance to H$\alpha$ images (with potentially greater opacity), their lateral contrast would be reduced due to an insensitivity to Doppler shifts. This reduced contrast could make them appear less distinct, particularly at smaller scales. It is worth noting that similarities between H$\alpha$ and ALMA features have been shown by \citet{2019ApJ...881...99M} and \citet{2021A&A...651A...6B}, who found correlated structures in simultaneous observations.
Although small dark fibrils are less commonly observed in (ALMA) millimeter continuum images compared to images taken in other chromospheric spectral lines, here we present identification of a long-lived fibril (lasting over the entire observation period), enabling wave studies to be directly undertaken along this structure. A thorough characteristic study of the same (unique) fibrillar structure has previously been conducted by \citet{2021ApJ...906...82C}, using spectral analysis provided by co-observations with the Interface Region Imaging Spectrograph (IRIS; \citealt{2014SoPh..289.2733D}), as well as a radiative magnetohydrodynamic 2.5D numerical model.

Our primary objective is to determine oscillation periods and phase speeds along the elongated fibrillar structure, with the ultimate goal of identifying the presence of MHD wave modes. By establishing a link between ALMA observations and specific MHD wave modes, we can refine our understanding of wave energy transport and dissipation within the complex chromospheric environment. Section~\ref{Observations} summarises the ALMA data analysed here. The wave analysis and results are provided in Sect.~\ref{Sect3}, and the concluding remarks are drawn in Sect.~\ref{discussion}.

\section{Observations}\label{Observations} 

The ALMA interferometric observations presented in this paper were taken on 22 April 2017 using C43-3 antenna configuration in Band 6 centered at 1.25~mm (i.e., 239~GHz) with project ID 2016.1.00050.S. 
This configuration used baselines from 14.6~m to 500~m with an elliptical beam with a median angular resolution of $0.84^{\prime\prime} \times 0.67^{\prime\prime}$ corresponding to $610 \times 487$~km$^{2}$ with a clockwise median inclination of 86.1$^{\circ}$ with respect to the solar north (i.e., the position angle). 
The observations targeted a plage region centered at heliographic coordinates N$11^{\circ}$E$17^{\circ}$, or at $(x,y)=(-260^{\prime\prime}, 265^{\prime\prime})$ in helioprojective coordinates.

The observations were carried out from 15:59 $-$ 16:38~UTC, producing 5 consequent scans (each 10 minutes) of the target region separated by four breaks (each 1.75 $-$ 2.25 minutes) for calibration, which were linearly interpolated before our wave analysis.

\begin{figure*}[]
 \centering
\includegraphics[width=0.95\textwidth]{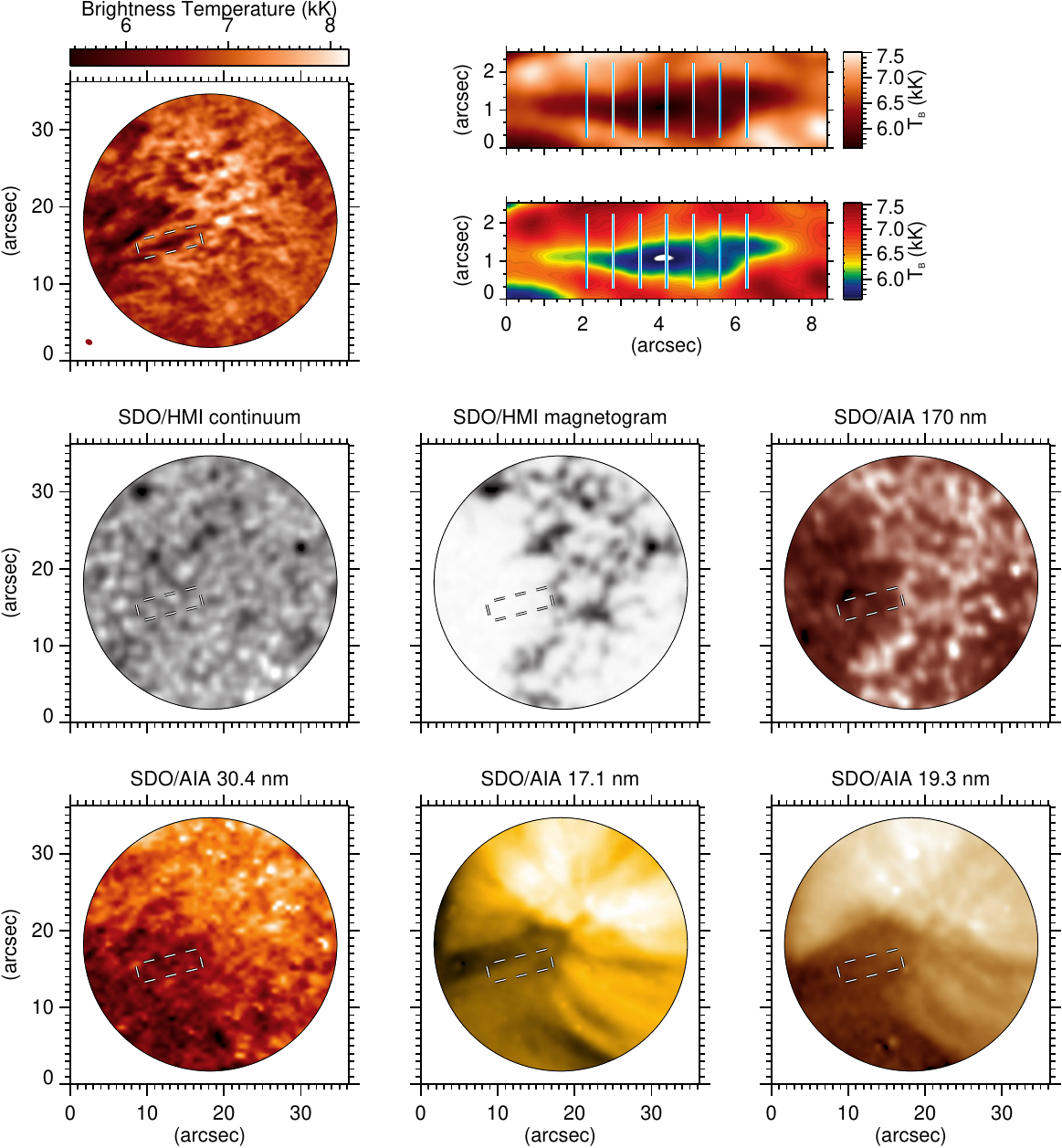}
 \caption[]{ Upper left panel: A brightness temperature ($\rm T_B$) map from the ALMA's Band 6 observation of the plage region at the start of time series. The ellipse on the bottom left corner of the panel represents the beam size of the observations and the white dashed rectangle outlines the location of the long-lived fibril of interest. Upper right panel: Zoomed-in view of the selected fibril (using two different colour table for better clarity) with the seven artificial slits placed perpendicular to the fibril axis for wave analysis as discussed in Sect. \ref{Sect3}. Middle and lower panels: co-aligned SDO/HMI continuum and magnetogram and SDO/AIA images at 170, 30.4, 17.1, and 19.3 nm. In the SDO/HMI magnetogram the line-of-sight photospheric magnetic fields ($\rm B_{los}$) is in range of -1116 < $\rm B_{los}$ (G) < 81. }
\label{ALMAimage}  
\end{figure*}

Data reduction and imaging were performed using the Solar ALMA Pipeline (SoAP; Szydlarski et al., in prep.), as detailed in \citet{2021ApJ...906...82C} and \citet{2022A&A...659A..31H}. This resulted in a time series of images with a high temporal resolution of 2~s and a spatial sampling (i.e., pixel size) of $0{\,}.{\!\!}''14$ per pixel. An image of ALMA Band 6 observations that is used in this study is illustrated in the left panel of Fig.~\ref{ALMAimage}.

Further details of these observations can be found in previous publications analyzing the same ALMA dataset. These include studies by \citet{2020A&A...634A..56D} on the temperature and microturbulence stratification in plage and quiet-Sun regions, \citet{2021ApJ...906...82C} and \citet{2021ApJ...906...83C} on an on-disk Type~{\sc ii} spicule (i.e., the fibrillar structure) and chromospheric plages, respectively, \citet{2021RSPTA.37900174J} on global p-modes (along with 9 other ALMA datasets), \citet{Narang22} on a detailed comparison of (global) oscillations power between ALMA and other UV channels co-observed with IRIS and the Solar Dynamic Observatory (SDO; \citealt{2012SoPh..275....3P}), and \citet{2023A&A...671A..69G} on MHD waves in small-scale bright features.

\section{Analysis and results}{\label{Sect3}}

The ALMA Band 6 observations analyzed in this study present small-scale bright features throughout the plage region, along with a few extended, fibrillar dark structures on the left-hand side of the field of view (FoV) (see Fig.~\ref{ALMAimage}). 
Using the same dataset, \citet{2023A&A...671A..69G} identified and detailed the properties of kink and sausage MHD wave modes in the small-scale bright structures. Field extrapolations of the photospheric magnetic fields (from the Helioseismic and Magnetic Imager, HMI \citealt{2012SoPh..275..229S}; see \citealt{2021RSPTA.37900174J}, Fig.~8) illustrate two main field topologies over the entire FoV: nearly vertical fields in the plage regions and nearly horizontal fields overarching a quiescent area (i.e., a neighboring internetwork region) in the areas covered by the extended and fibrillar dark structures.

In our Band 6 data, we do not observe many individual dark fibrils. Instead, we find a few extended dark structures, likely due to unresolved individual fibrils. However, we identified only one long-lived fibril (along with a few short-lived ones). Thus, we study here oscillatory signatures and their properties along the detected long-lived fibril.

\subsection{Identification of oscillatory signals}\label{Trace-fibrils}

To study waves and oscillations in the observed long-lived fibril, we track the dark elongated structure present in all frames, corresponding to a lifetime of 1568~s.
We use the identification and tracking approach as previously introduced by \citet{2017ApJS..229....7G,2017ApJS..229....6G}. This method uses an unsharp mask algorithm and an adaptive histogram equalization procedure to enhance the intensity contrast between neighboring pixels and within the image (only for detection purposes). Using this method, we identified the long-lived fibril that persisted throughout the observed period. The spatial location of the fibril is marked by the dashed-line rectangle in the first frame of the time series, in the left panel of Fig.~\ref{ALMAimage}.

To quantify oscillatory signals along the fibril, we placed seven artificial slits perpendicular to the fibril axis, spaced 5 pixels apart (corresponding to 510 km on the solar disc). The upper right panel of Fig.~\ref{ALMAimage} shows the fibril in the first frame, with the seven slits marked as vertical lines. The slits were positioned in the part of the fibril visible throughout the time series, as the fibril's length varied over time. The lower panels of Fig.~\ref{ALMAimage} show SDO/HMI continuum and magnetogram and SDO atmospheric Imaging Assembly (AIA; \citealt{2012SoPh..275...17L}) images at various wavelengths. We also examined H$\alpha$ images from the Global Oscillation Network Group (GONG; \citealt{2011SPD....42.1745H}), however, due to their relatively low spatial resolution, no corresponding features were visible within the ALMA field of view. While no clear counterpart was identified in the relatively low-resolution GONG H$\alpha$ images, we anticipate a correspondence with higher-resolution H$\alpha$ observations. \citet{2019ApJ...881...99M} found a strong correlation between ALMA 3 mm observations and the spatial structure of chromospheric features seen in high-resolution H$\alpha$ line-core images. Based on their findings, we expect this fibril to have a corresponding counterpart visible in high-resolution H$\alpha$ data. Similarly, we expect corresponding features to be present in EUV observations, particularly in AIA channels sensitive to transition region and coronal temperatures, although high-resolution observations are necessary to resolve these relatively small structures. The co-aligned SDO/AIA images in the lower panel of Fig.\ref{ALMAimage} shows that the fibril is located near a larger EUV loop (fibrillar structure) in the corona. That is less clear in the noisy 30.4 nm images. However, we note that larger-scale structures would likely show more similarities between ALMA and 30.4~nm images, as shown by \citet{2018A&A...613A..17B, 2020A&A...635A..71W}. Additionally, as is evident from the SDO images, the fibril lies above an internetwork region characterized by a magnetic canopy structure at chromospheric and coronal heights. \citet{2021RSPTA.37900174J}, which calculated the approximate magnetic topology of these observations (from field extrapolations of the HMI observations), confirmed the predominantly horizontal magnetic field configuration in this region.

We measured the brightness temperature in all pixels along each slit and determined the fibril's position at each location by finding the centroid of the slit using Gaussian fitting. Additionally, we calculated the brightness temperature at the centroid (representing the fibril's intensity) and the full width at half maximum (FWHM) of the Gaussian fit (representing the fibril's width) at each slit location.

\subsection{Wavelet analysis}

To characterize the wave properties of brightness temperature, width, and transverse displacement, we performed a wavelet analysis. Wavelet analysis localizes spectral power in both time and frequency by decomposing the signal into time-localized wavelets, revealing how the frequency content of the signal changes over time \citep{Daubechies90,Torrence98}. We identified and removed the Cone of Influence (CoI), the unreliable areas in the time-frequency space subject to edge effects, from further analyses.
We note that all signals (corresponding to brightness temperature, width, and position) were detrended by using a linear fit and then apodised using a Tukey window prior to the wavelet analyses. Furthermore, periods larger than 1000~s (i.e., smaller than 1~mHz) were subtracted from all signals by means of wavelet filtering. Such long periods are likely related to slow evolution of the magnetic structure, rather than wave signatures.

We calculated the wavelet power spectra of oscillations in all three parameters across the seven slits along the fibril using a Morlet function. This yielded the periods (or frequencies) of the fluctuations and their variations over time for each parameter. Additionally, we calculated the wavelet cross-power spectra between consecutive slits for each parameter, obtaining the phase relationships between the oscillations and, thus, the phase speeds of the waves propagating along the fibril.

\begin{figure*}
  \centering
  \makebox[1\textwidth]{{Brightness Temperature}}\par
  \subfloat{\includegraphics[width=.32\textwidth]{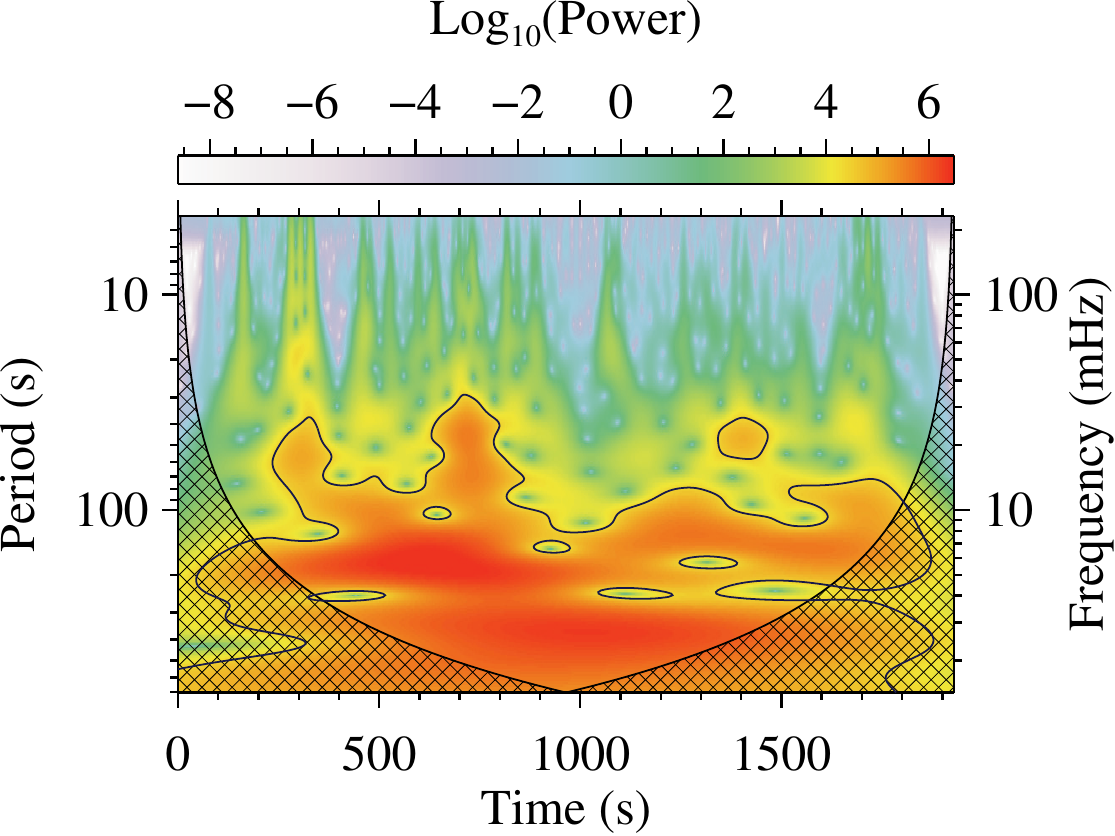}}\hspace{0.2cm}
  \subfloat{\includegraphics[width=.32\textwidth]{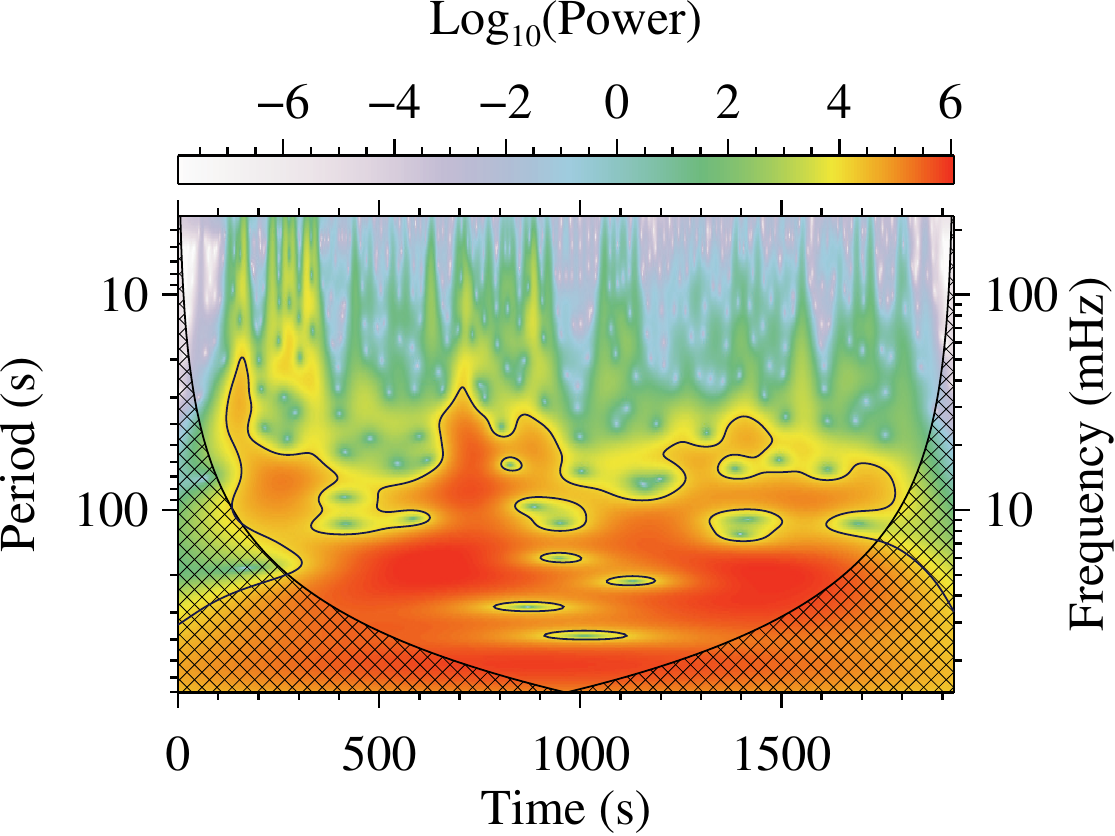}}\hspace{0.2cm}
  \subfloat{\includegraphics[width=.32\textwidth]{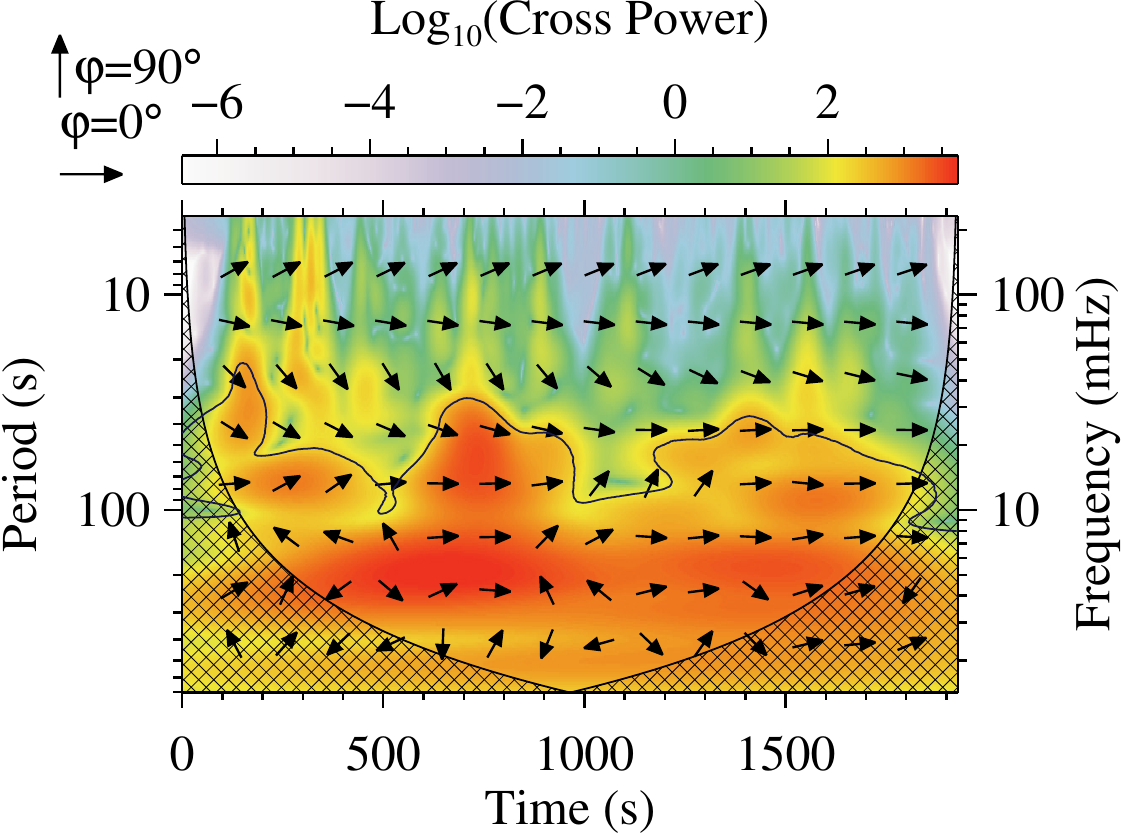}}\par
  \vspace{0.3cm}
  \makebox[1\textwidth]{{Horizontal Displacement}}\par
  \subfloat{\includegraphics[width=.32\textwidth]{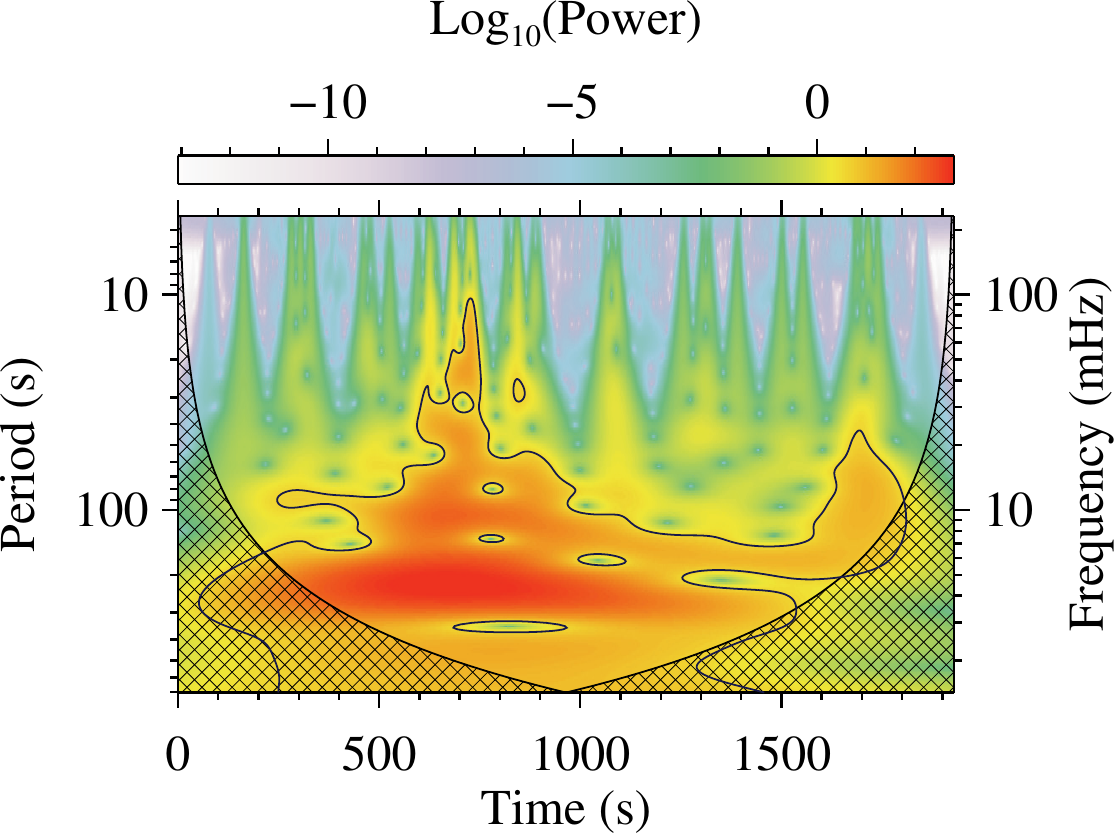}}\hspace{0.2cm}
  \subfloat{\includegraphics[width=.32\textwidth]{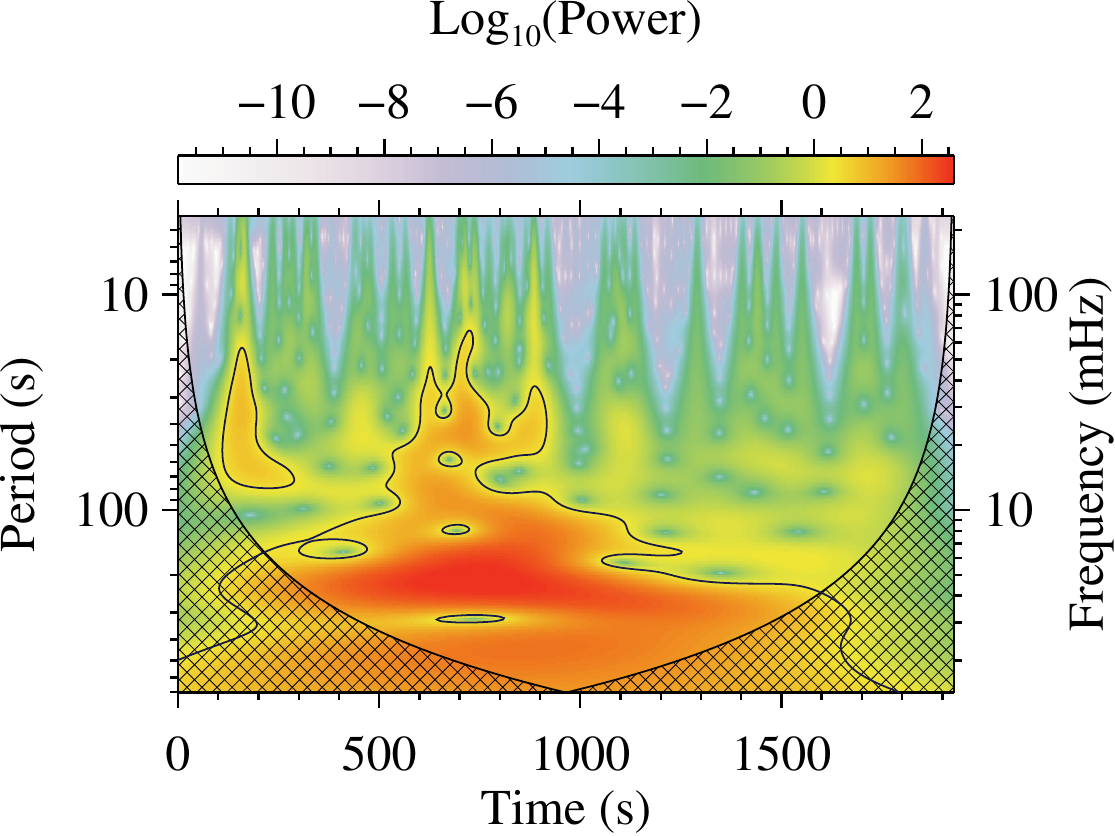}}\hspace{0.2cm}
  \subfloat{\includegraphics[width=.32\textwidth]{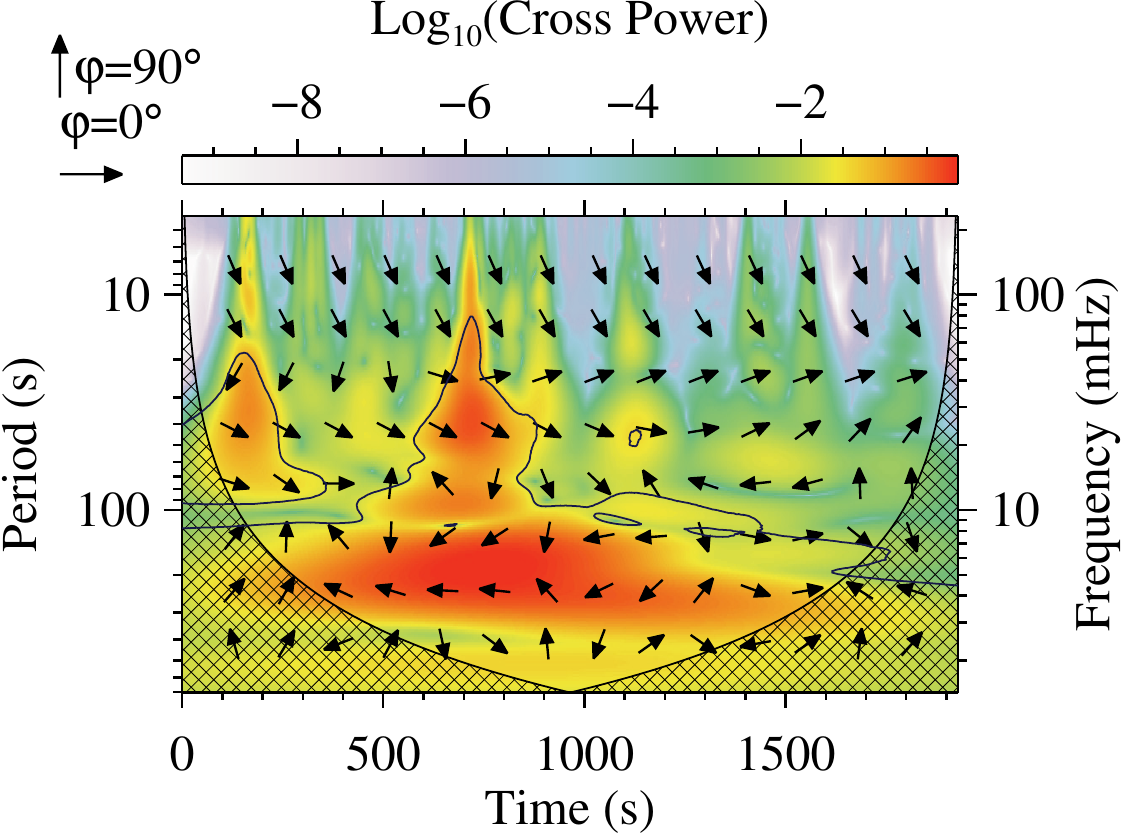}}\par
  \vspace{0.3cm} 
  \makebox[1\textwidth]{{Width}}\par
  \subfloat{\includegraphics[width=.32\textwidth]{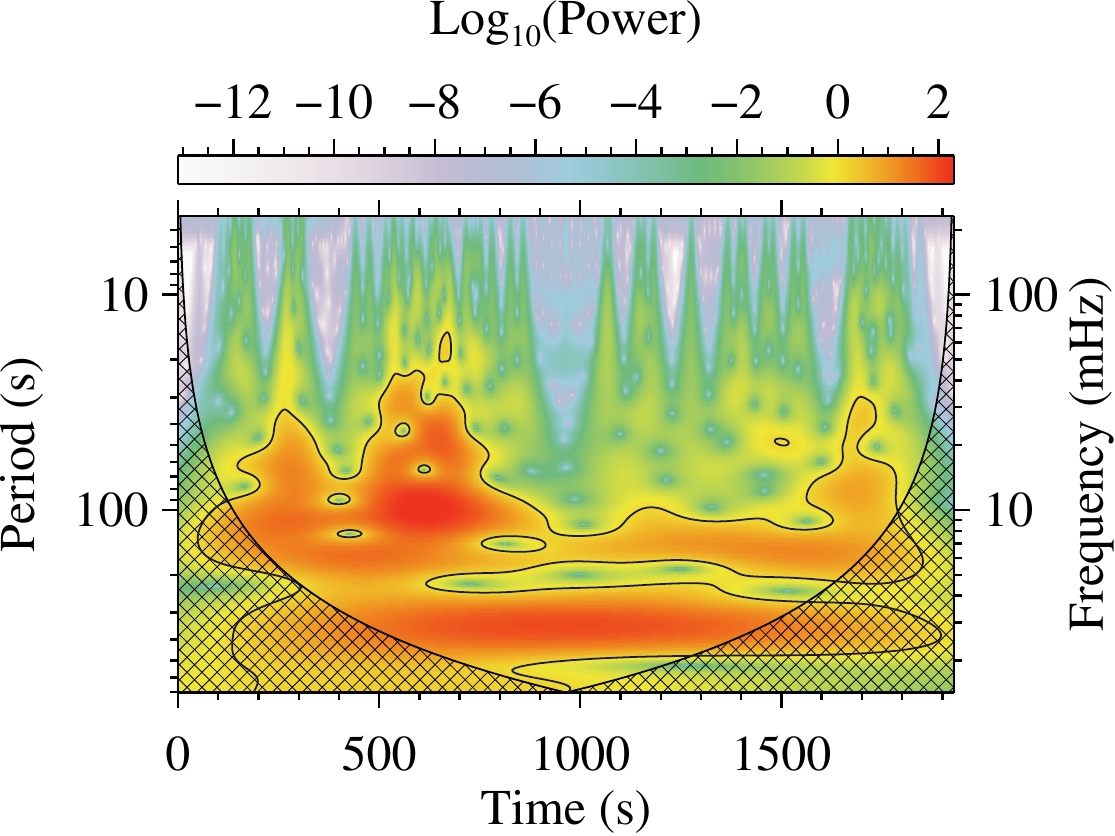}}\hspace{0.2cm}
  \subfloat{\includegraphics[width=.32\textwidth]{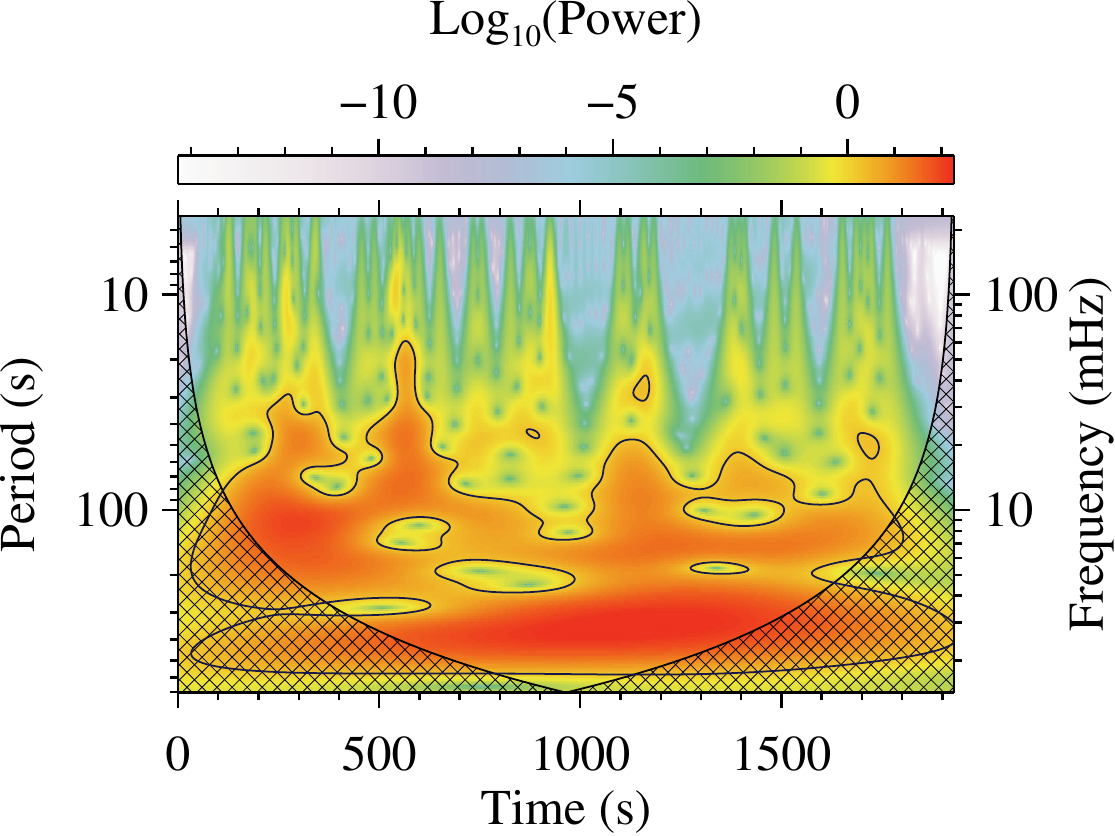}}\hspace{0.2cm}
  \subfloat{\includegraphics[width=.32\textwidth]{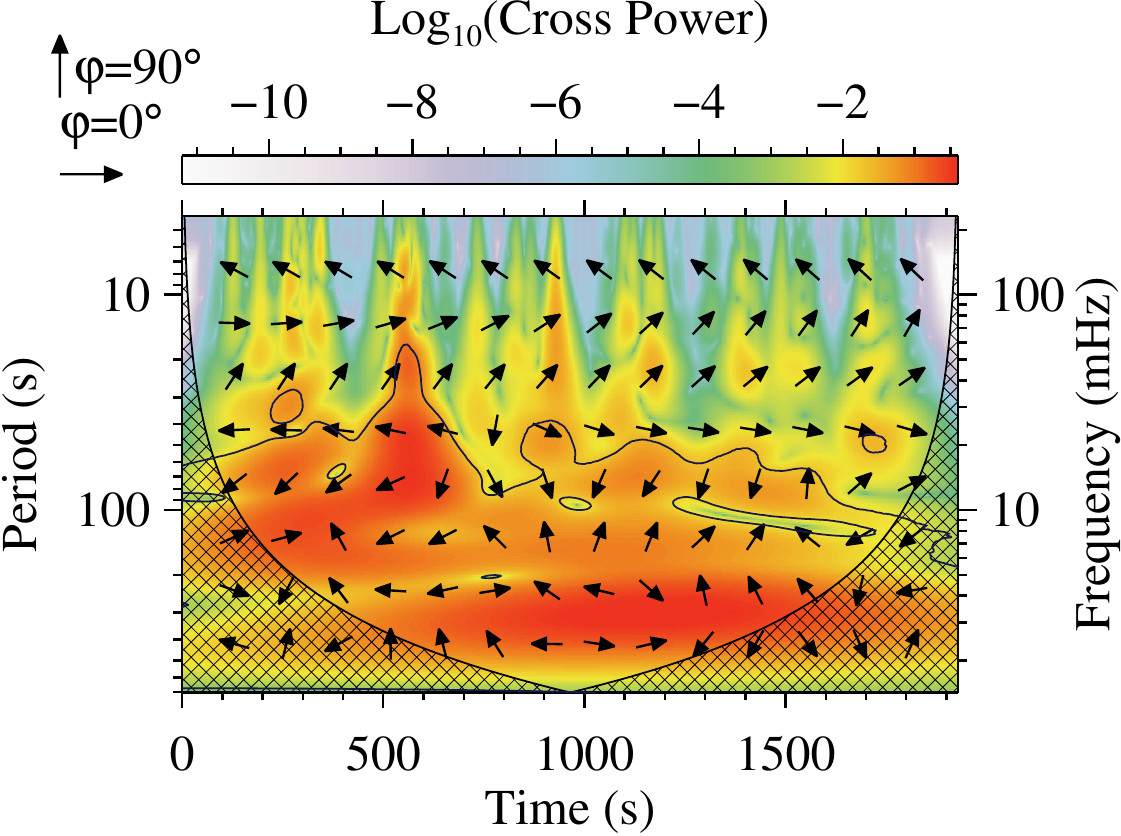}}
  \caption{Wavelet power spectra (left and middle columns) and wavelet cross-power spectra (right column) for oscillations in brightness temperature (top row), horizontal displacement (middle row), and width (bottom row) at two consecutive slits along the fibril. The cross-hatched areas indicate the wavelet's cone of influence. In the cross-power spectra, arrows represent the phase relationship between oscillations at the two slits: rightward arrows indicate in-phase oscillations ($0^\circ$), leftward arrows indicate anti-phase oscillations ($180^\circ$), and downward arrows indicate the second slit leading the first by $90^\circ$. Black contours in all panels mark the $95\%$ confidence level. }
  \label{wavelet}
\end{figure*}

Figure \ref{wavelet} illustrates examples of the wavelet power spectra. The left and middle columns show the power spectra of the oscillations of the three parameters at two consecutive slits (third and forth slits) and the right column presents their corresponding wavelet cross-power spectra. The CoI regions are marked as cross-hatched areas and the black contours identify the $95\%$ confidence levels. The arrows on the wavelet cross-power spectra indicate the relative phase relationship between oscillations in consecutive slits. Arrows pointing to the right represent in-phase oscillations ($\phi=0^\circ$), arrows pointing to the left represent anti-phase oscillations ($\phi=180^\circ$), and arrows pointing down represent a $90^\circ$ phase shift, where the second oscillation leads the first, indicating wave propagation from left to right along the fibril. In-phase and anti-phase relationships are indicative of standing waves.

\begin{figure*}
    \centering
    \includegraphics[width=1.0\textwidth]{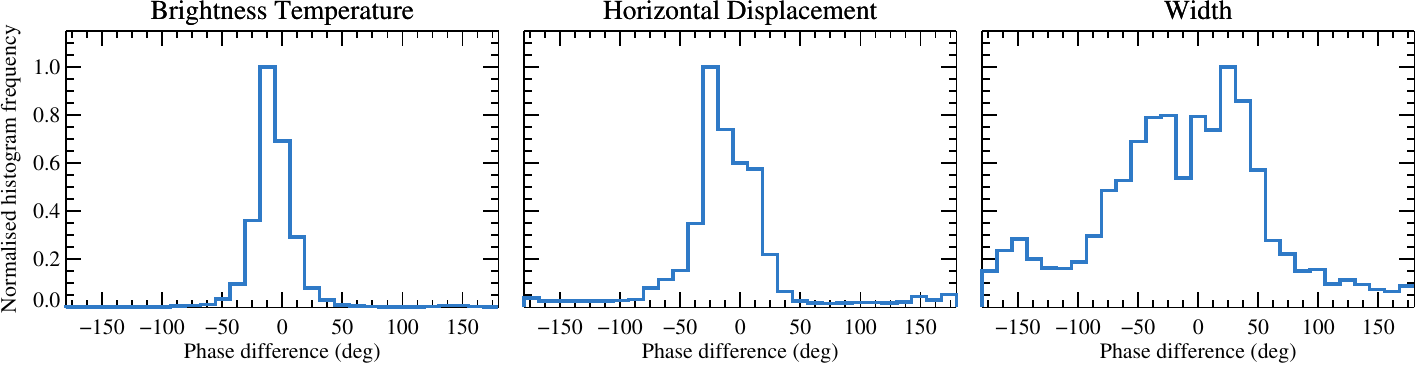}
  \caption{Normalized distributions of phase differences between oscillations in brightness temperature (left), horizontal displacement (middle), and width (right) at pairs of consecutive slits along the fibril. }
  \label{Hist-phase}
\end{figure*}

As shown in the right column of Fig. \ref{wavelet}, all three parameters (brightness temperature, horizontal displacement, and width) exhibit a combination of standing and propagating waves. The normalized phase-lag distributions (Fig. \ref{Hist-phase}) reveal a dominance of in-phase and anti-phase relationships, particularly for brightness temperature and horizontal displacement, suggesting a prevalence of standing waves. However, the presence of numerous other phase values indicates that propagating waves are also present. We note that some of the abrupt changes in phase angles could be due to noise or spurious signals.

We also calculated the wavelet cross-power spectra between pairs of parameters (brightness temperature $\&$ position, brightness temperature $\&$ width, and position $\&$ width) at each slit to determine the phase correlation of oscillations within the same slit. However, interpreting such relationships is complex due to the potential superposition of multiple MHD wave modes, which current theoretical models are not yet sophisticated enough to fully disentangle, as discussed by \citet{2024A&A...688A...2J}. Preliminary analysis suggested that brightness temperature and position oscillations primarily exhibited in-phase and anti-phase relationships, while brightness temperature and width oscillations showed a more isotropic distribution of phase differences with slight peaks at $0^\circ$ and $\pm180^\circ$. Position and width oscillations displayed diverse phase angles, but in-phase and anti-phase relationships still dominate, though less strongly than for brightness temperature and position.

\subsubsection{Wave periods}

\begin{figure*}
    \centering
    \includegraphics[width=1.0\textwidth]{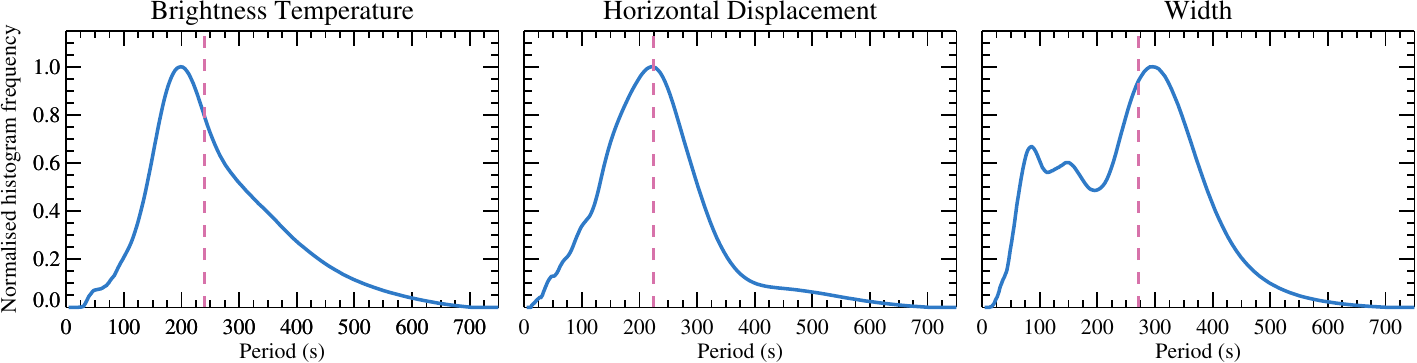}
    \caption{Normalized distributions of oscillation periods in brightness temperature, horizontal displacement, and width derived from wavelet cross-power spectra. The vertical dashed lines identify the median of each distribution.}
    \label{Hist-period}
\end{figure*}

Figure \ref{Hist-period} presents the normalized distributions of the wave periods for brightness temperature, horizontal displacement, and width along the fibril. These distributions are constructed by considering all power-weighted periods within the $95\%$ confidence levels and outside the CoI of the wavelet cross-power spectra calculated for each slit. The final distribution is then normalized by its maximum value to facilitate comparison.

The distributions for all three parameters exhibit a broad range of periods, extending from approximately 15 to 700~s. The brightness-temperature and horizontal-displacement histograms display single-peak distributions with distinct peaks at 200 and 225~s, respectively. The width distribution shows a primary peak at 300~s and two less pronounced peaks around 80 and 150~s.

Table \ref{periods} summarizes the range, mean, median, and standard deviation of the oscillation periods for each parameter, as determined from the wavelet analysis.

\begin{table}[]
\caption{Range, mean, median, and standard deviation of oscillation periods (in seconds) for brightness temperature, horizontal displacement, and width calculated across all slits along the fibril.}
  \centering
  \setlength{\tabcolsep}{4.0pt}
\begin{tabular}{lllll}
\hline
Parameter (s) & Range & Mean & Median & Std. Dev. \\
\hline
Brightness temperature & 21-692 & 264 & 241 & 114 \\
Horizontal displacement & 11-692 & 233 & 224 & 102 \\
Width & 13-692 & 260 & 272 & 118 \\
\hline
\end{tabular}
\label{periods}
\end{table}

\subsubsection{Phase speeds}

The phase lag between two points in a traveling wave represents how much one wave has advanced relative to the other, directly translating to a travel time between those points. In a standing wave, a $\pm180^\circ$ phase-lag signifies a special relationship of persistent anti-phase oscillation, rather than a measure of propagation. However, standing waves can result from the superposition (interference) of two waves traveling in opposite directions, each with its own phase speed.

To quantify wave propagation, we calculate the propagation time ($\tau$) between pairs of consecutive slits along the fibril using:
\begin{equation}
\tau = \frac{\phi P}{2\pi},
\label{Tlag}
\end{equation}
where $\phi$ is the phase angle (excluding $\pm180^\circ$) and $P$ is the wave period, both obtained from the wavelet analysis. In this calculation, we also consider only phase differences that fall within the $95\%$ confidence levels and outside the CoI of the wavelet cross-power spectra. These phase differences are further weighted by power, emphasizing the significance of correlation strengths in the time-frequency domain.

\begin{figure*}
    \centering
    \includegraphics[width=1.0\textwidth]{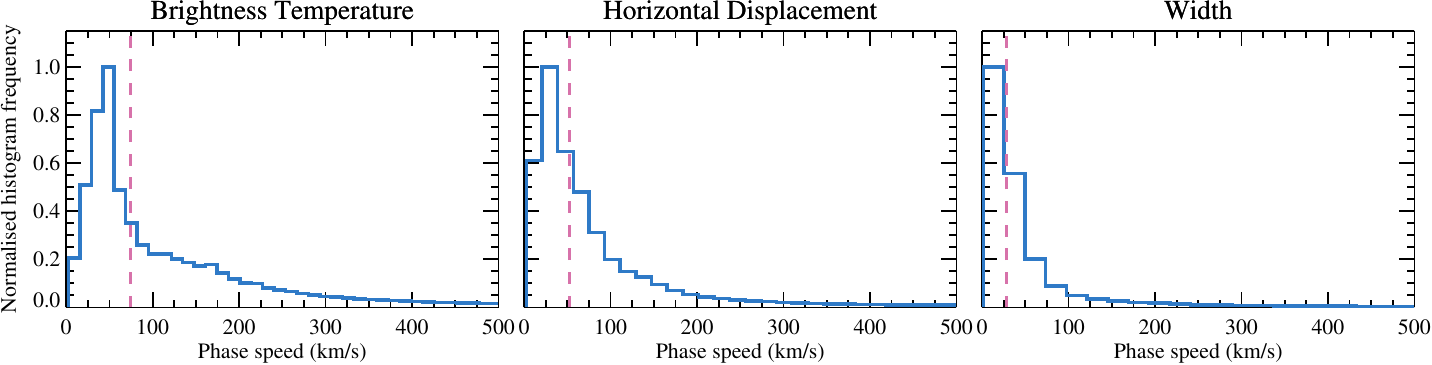}
 \caption{Normalized distributions of absolute phase speeds for oscillations in brightness temperature, horizontal displacement, and width along the fibril. The vertical dashed lines mark the median of each distribution.}
  \label{Hist-vel}
\end{figure*}

Knowing the fixed distance between slits (510~km) and the calculated time lags, we determine the phase speed (or propagation speed) for brightness temperature, width, and position oscillations. Figure \ref{Hist-vel} shows distributions of the absolute values of phase velocities, that include both leftward-propagating and rightward-propagating waves. The co-existence of these oppositely propagating waves in the same structure may suggest that at least some of the observed standing wave patterns (those with $0^\circ$ phase lags) may result from their superposition.

Table \ref{phase-speed} summarizes the mean, median, and standard deviation of the propagation speeds, along with the occurrence rates of leftward and rightward propagation.

\begin{table}[]
\caption{Mean, median, and standard deviation of phase speeds and their absolute values (in km/s) for oscillations in brightness temperature, horizontal displacement, and width along the selected fibril. The last two columns show the percentage of rightward-propagating (positive velocities) and leftward-propagating (negative velocities) waves.}
  \centering
  \setlength{\tabcolsep}{1.5pt}
\begin{tabular}{l|ccc|cc}
\hline
Parameter & Mean & Median & Std. Dev. & Right. & Left. \\
\hline
Brightness temperature  & -29  & -44 & 250 & 28$\%$ & 72$\%$ \\ 
absolute values & 148 & 74 & 204 &  & \\
\hline
Horizontal displacement & -12 & -27 & 222 & 32$\%$ & 68$\%$ \\
absolute values & 103 & 52 & 197 &  & \\
\hline
Width & -3 & -6 & 268 & 46$\%$ & 54$\%$ \\
absolute values & 86 & 28 & 254 &  & \\
\hline
\end{tabular}
\label{phase-speed}
\end{table}

\section{Discussion and Conclusions}\label{discussion}

Using high-cadence (2~s) ALMA Band 6 observations, we investigated the presence of MHD waves in a long-lasting ($\approx $1600~s) dark fibrillar structure, which represent strong magnetic field bundles in the upper chromosphere. By placing multiple equally-spaced (510~km) artificial slits perpendicular to the fibril axis, we examined oscillations in brightness temperature, transverse displacement, and width along the length of each slit.

Our analysis revealed a combination of both standing and propagating waves in all three parameters, with oscillation periods ranging from approximately 15 to 700~s. The period distributions of brightness temperature and transverse displacement exhibited single-peaked histograms, with median and standard deviations of $240\pm114$~s, $225\pm102$~s, respectively. The width distribution showed a primary peak at 300~s and two less pronounced secondary peaks around 80 and 150~s, with a period of $272\pm118$~s as the median and standard deviation of the distribution.

The propagating waves exhibited mean absolute phase velocities (with median and standard deviation) of $74\pm204$~km/s for brightness temperature, $52\pm197$~km/s for transverse displacement, and $28\pm254$~km/s for width. The transverse oscillations likely represent the presence of MHD kink modes \citep{1982SoPh...75....3S}, while the size fluctuations (width oscillations) likely indicate sausage modes \citep{1981SoPh...69...39R, 2013A&A...555A..75M}.

Our findings are consistent with previous studies of MHD waves in chromospheric structures observed with ALMA. \citet{2021RSPTA.37900184G} reported similar oscillatory behavior in bright point-like features, identifying fast sausage modes with average oscillation periods of $90\pm22$~s for brightness temperature and $110\pm12$~s for size. These periods are shorter than those found in our dark fibril, potentially due to variations in magnetic field strength and geometry between the two types of structures.

Subsequent studies using ALMA have also focused on small-scale bright features, interpreted as cross-sections of magnetic flux tubes (manifesting as fibrils when oriented horizontally). \citet{2022A&A...665L...2G} detailed the propagation of transverse kink waves in these structures using ALMA sub-bands, while \citet{2023A&A...671A..69G} analyzed MHD wave modes in a larger sample. They found evidence of transverse (kink) waves with average amplitude velocities of 2-4.3~km/s and compressible sausage modes, with average oscillation periods of 70-110~s for brightness temperature, 61-100 s for size, and 57-80~s for horizontal velocity. These values of periods and velocities are smaller than those observed in our dark fibril, likely because these small bright features form at lower atmospheric heights.

Comparing with studies of MHD waves in fibrillar structures observed in other diagnostics, both kink and sausage modes were detected in bright slender Ca~{\sc ii}~H fibrils (located in the low-to-mid chromosphere) from high-resolution {\sc Sunrise} observations \citep{2017ApJS..229....2S}, with periods on the order of 83 and 35 s, and phase speeds of $9\pm14$ and 11-15~km/s respectively \citep{2017ApJS..229....9J,2017ApJS..229....7G}. Longer periods (232 and 197 s) and larger propagating speeds (80 and 67 km/s) were reported by \citet{2012NatCo...3.1315M} for higher chromospheric (dark) fibrillar structures, comparable with those we found in the present work.

While relatively short periods (shorter than, e.g., 100~s) have been more frequently observed in the low chromosphere, they are less common in the upper chromosphere \citep[see also,][]{2007ApJ...655..624D,2011ApJ...739...92P,2012ApJ...750...51K,2013ApJ...768...17M,2014ApJ...784...29M,2021RSPTA.37900183M}, where our ALMA fibril is likely located \citep{2021ApJ...906...82C}. Such short-period waves could possibly be dissipated through the chromospheric heights before reaching the upper layers.

Leveraging both observational data and numerical simulations, \citet{2012ApJ...744L...5J} suggested that mode conversions in the lower solar atmosphere can be a main driver of the MHD kink and sausage modes observed in chromospheric fibrillar structures (in on-disk Type~{\sc i} spicules). This mechanism could also play a role in the excitation of the waves we observe in the upper chromospheric fibril.

Theoretical studies have shown that asymmetry in waveguides, such as the dark fibril studied here, can influence MHD wave properties, including the coupling of different wave modes, shifts in phase speeds, and modifications in damping rates \citep{2017SoPh..292...35A, 2018ApJ...855...90A, Erdelyi2024}. Furthermore, irregularities within these waveguides can significantly alter the properties and characteristics of resonant modes, particularly those of higher MHD wave modes \citep{2021RSPTA.37900181A, 2022ApJ...927..201A}. While our current analysis primarily focuses on identifying kink and sausage modes, future investigations incorporating the effects of waveguide asymmetry could offer deeper insights into these effects and their implications for wave dynamics observed in these dark fibrils. Such studies would require detailed modeling of the fibril's cross-sectional shape and density structure, which could be achieved by combining high-resolution observations with advanced MHD wave modeling techniques.

In conclusion, our study provides a first evidence for the ubiquitous presence of MHD waves in dark fibrils observed by ALMA in Band 6. The detection of both standing and propagating waves, alongside the identification of potential kink and sausage modes, underscores the complex wave dynamics within these structures. The distinct wave properties observed here, compared to other chromospheric features, highlight the importance of considering factors like magnetic field strength and geometry when interpreting wave phenomena. Importantly, this work demonstrates ALMA's capability to effectively sample such dynamic dark fibrillar structures, despite previous doubts. Future statistical studies of these fibrils using higher-resolution ALMA observations across multiple bands promise to further enhance our understanding of MHD wave behavior in the chromosphere.

\begin{acknowledgements}

This work is supported by the SolarALMA project, which received funding from the European Research Council (ERC) under the European Union's Horizon 2020 research and innovation programme (grant agreement No. 682462), and by the ESGC project (project No. 335497) funded by the the Research Council of Norway, and by the Research Council of Norway through its Centres of Excellence scheme, project number 262622.
S.J. gratefully acknowledges support from the Rosseland Centre for Solar Physics (RoCS), University of Oslo, Norway. 
D.B.J. acknowledges support from the Leverhulme Trust via the Research Project Grant RPG-2019-371, as well as the UK Space Agency via the National Space Technology Programme (grant SSc-009). D.B.J. also wishes to thank the UK Science and Technology Facilities Council (STFC) for the consolidated grants ST/T00021X/1 and ST/X000923/1. 
R.G. acknowledges the support by Fundação para a Ciência e a Tecnologia (FCT) through the research grants UIDB/04434/2020 and UIDP/04434/2020.
This paper makes use of the following ALMA data: ADS/JAO.ALMA\#2016.1.00050.S. ALMA is a partnership of ESO (representing its member states), NSF (USA) and NINS (Japan), together with NRC (Canada), MOST and ASIAA (Taiwan), and KASI (Republic of Korea), in cooperation with the Republic of Chile. The Joint ALMA Observatory is operated by ESO, AUI/NRAO and NAOJ. We are grateful to the many colleagues who contributed to developing the Solar observing modes for ALMA and for support from the ALMA regional centres.
We wish to acknowledge scientific discussions with the Waves in the Lower Solar Atmosphere (WaLSA; \href{https://WaLSA.team}{www.WaLSA.team}) team, which has been supported by the Research Council of Norway (project no. 262622), The Royal Society (award no. Hooke18b/SCTM; \citealt{2021RSPTA.37900169J}), and the International Space Science Institute (ISSI Team 502).

\end{acknowledgements}


\bibliographystyle{aa} 
\bibliography{references}   

\end{document}